\documentclass[aps,prl,twocolumn,groupedaddress,showpacs,tightenlines]{revtex4}
\begin{document}

\title{Variation of Fine Structure Constant from Non-Universal Gravity}

\author{Johann Rafelski}
 \affiliation{Physics Department, University of Arizona, Tucson, Arizona 85721}

\date{\small August 28, 2002}

\begin{abstract}
We relate the reported variation in the value of the fine structure constant 
to a possible non-universality of the gravitational interaction with 
respect to different particle generations.
\end{abstract}
\maketitle
Recent measurement  of the fine structure constant
$\alpha$ in the remote, yet on cosmological scale recent past 
yielded a $4\sigma$ deviation  $\Delta\alpha/\alpha=-(0.72\pm0.18) 10^{-5}$
from laboratory value within the red shift
domain $z\simeq $0.6--2 \cite{Web01}.
This poses the question about the origin of 
such an effect \cite{Ban02,Dav02}, identified as either 
temporal (general presumption) or spatial 
variation. Since the experimental effect 
which is observed is seen within a scheme 
of a systematic change in absorption 
lines by nebula  in line of sight  of 
quasar illumination of laboratory sensors, 
the effect could be originating in properties 
of the nebula. Such spatial interpretation 
of the data analysis is more compatible with 
other limits on variation of $\alpha$ \cite{Web01}. 
 
As recent flurry of activity has shown that 
it is quite difficult to  find a   self consistent 
interpretation of this experimental result. We propose here a 
quite different approach:\\
a) We  interpret   the change  in $\alpha$ as being due to
fermion masses $m_f$ which enter vacuum polarization loops.\\ 
b) We will argue that the effect cannot involve electrons, thus
 only the heavy fermions  such as the muon contribute.\\
c) Assuming that  heavy fermion masses have varied
by about -0.15--0.45\% this experimental observation can be understood.\\ 
d) There is no consistent interpretation assuming 
time variation of $m_f$ (or $\alpha$) since there is no 
sensitivity to the time evolution of the late Universe we live in.\\
e) When the effect is associated with spatially localized 
gravitational mass  it requires  non-universal gravitational coupling.\\
f) We show how a theory of gravity is formulated in principle which 
combines different ratios of gravitational to inertial mass for the
three fermion generations. This is not in contradiction with 
the E\"otv\"os experimental evidence for universality of the ratio of
inertial to gravitational mass, and all  principal results
of Einstein gravity and cosmology are retained.\\
g) The 1000-fold enhanced strength of gravity for 2nd and/or 3rd 
generation, which is required to understand $\Delta\alpha/\alpha$,
also explains the dark matter in terms of the more strongly
gravitating cosmic background 2nd and/or 3rd generation
neutrinos.\\
h) We do not present here a theoretical model how a gauge universal,
gravity non-universal, theory could arise.\\

\vskip 0.3cm
{\bf Conventional theoretical framework:}
Within the current paradigm of a  unified 
structure of all interactions, the value of $\alpha$, 
the electromagnetic coupling in  the infrared large distance 
$r\to \infty$ limit is  not fundamental, the determining
value is the strength of the universal interaction at the unification 
scale, which in case of `Grand Unification' is
 $M_{\rm GUT}\simeq 3\cdot 10^{16}$ GeV. As energy scale 
diminishes, broken  symmetries emerge and individual
 interactions separate, and their strength evolves
with the scale.  Electromagnetic (EM) interaction 
 emerges when the electro-weak (EW) interactions 
separate at the scale of 
$\cal O$(100 GeV), and  where~\cite{Bur01,Jeg01},  
\begin{equation}\label{alfaZ}
\alpha(M_Z=91.2 \mbox{GeV})= \frac 1{128.936(46)}. 
\end{equation}

Unification of interactions  poses a problem for an apparent time 
variation of $\alpha$, since it implies a much larger
associated  variation in strong interaction coupling 
parameters, implicating significant changes in quantities such as 
the nucleon mass and nuclear interactions \cite{Lan02,Cal02}, 
and indeed leading to possible and probable 
disagreement with latest limits on cosmological variation 
of e.g. the proton to electron mass \cite{Lev02}.  
In addition  we need to resolve the 
question, how can sensitivity to the structure
of the universe arise at a relatively late stage of its evolution,
or if spatial interpretation is preferred, in localized nebula.

Introduction of relatively large compactified extra
dimensions reduces the unification scale considerably, 
to the TeV range \cite{Die98}.  Association of 
the time dependence of $\alpha$ or as we argue here $m_f$
 with large extra dimensions 
has the inherent flaw that these dimensions do not evolve beyond the 
early, formative period of the Universe \cite{Die02}.

The case is quite strong today,
in view of the many ramifications \cite{Oli02} of a 
time-changing strong interaction
strength accompanying the change in $\alpha$, 
that it is necessary  to  search for other 
possible sources of, on cosmological scale `nearby', 
time, or spatial, dependence of  $\alpha$.
Regarding the forthcoming discussion of the three particle generations
we recall that it is not understood today how the 
masses of the `heavy' fermion 
families are generated from a unified theory involving radically
different scales  (hierarchy problem). 
It is expected that in a unified theory which comprises
the quantum gravity, not only the interactions, but also 
the three particle family generations
will be naturally explained. 

The fundamental generation which 
is at the origin of the matter around us, also provides
the measuring sticks in the Universe. The unified theory than 
explains the mass  ratios such as 
the muon to electron mass is $m_\mu/m_e=206.77$. 
Furthermore, if electron mass
(along with other lepton masses)  were to  vary,   in
its units the nuclear world around us would evolve, and this
is not safe from  contradictions with precision experiments. One can 
more generally argue considering the impact of masses on
fine-tuned properties of known interactions that 
all  the fundamental generation masses ($m_e,m_{\nu_e}, m_d,m_u$) 
must not vary. Thus only variations in the masses
of the second  ($m_\mu, m_{\nu_\mu}, m_s, m_c$) and third 
($m_\tau, m_{\nu_\tau}, m_b, m_t$) generation could be considered  
to be at the origin of the variation of $\alpha$.

Although we make this argument with time dependence of $\alpha$
in mind, it applies equally for the case we explore below of
a spatial variation of $\alpha$. When we speak of variation 
of mass, we also include in this the gravitational mass defect.

\vskip 0.3cm
{\bf Relation of $\alpha$ to fermion masses:}
The scale evolution of the EM
coupling strength $\alpha\equiv\alpha(M=0)= 1/137.03599976(50)$, 
arises from the interaction of  photons with the virtual 
fluctuations of charged fermi particle pairs of mass $m_f$ with
 mass thresholds within the 
range of energy scale considered. 
Quantum electrodynamics allows to evaluate this vacuum polarization effect
for leptons with great precision. For our consideration 
it is sufficient to recall the
lowest order in electromagnetic interaction, and the large $M/m$ behavior: 
\begin{eqnarray}\label{alpharun}
\Delta\alpha^{-1}&\equiv&\frac1{\alpha(M)}-\frac1{\alpha(0)}\\
\nonumber &=&
  -\frac 2{3\pi}\sum_fQ^2_f N_f\left(\ln \frac{M}{m_f}
       -\frac 56 + {\cal O}(\frac{m_f}{M})\right).
\end{eqnarray} 
Here $Q_f$ is the charge of the fermion $f$ and $N_f$ is the 
internal degeneracy, such as color of quarks (=3). $M$ is the scale
at which the interaction is stable as function of time, and/or
environment.  In our present 
considerations we assume that all the variation occurs within
the energy scale  domain below $M_Z$, in which EM emerged as separate
interaction. This does not restrict the validity of our principal
argument.

 Eq.\,(\ref{alpharun})
can be applied to evaluate the contribution of the 
electron, muon and tau  $f=e,\mu,\tau$ leptons, 
and to estimate the effect 
of strongly interaction quarks, $u,d,s,c,b$
to the `running' of $\alpha(M)$. 
The contribution of quarks which also interact strongly, 
is more precisely determined
combining dispersion methods with experimental 
cross section results \cite{Bur01},
rather than through the use
of Eq.\,(\ref{alpharun}). The lepton contribution is
$\Delta\alpha^{-1}_l=4.3164$ 
(including a very small higher order effect)
while the hadron contribution is \cite{Jeg01} 
$\Delta\alpha^{-1}_h=3.823\pm0.054 $. 
In comparison, the reported cosmological variation 
is \cite{Web01} can be expressed as
$$\delta(1/\alpha)=\frac{-1}{\alpha}\frac{\delta\alpha}{\alpha}
\simeq 0.001,$$ 
which is 50 times smaller than the uncertainty
remaining in the understanding how the quark degrees 
of freedom influence the
evolution of EM interaction strength, see Eq.\,(\ref{alfaZ}).  

From Eq.\,(\ref{alpharun}) we obtain 
how $\alpha(0)$  responds to the  variation of fermion  masses,
assuming that $\alpha(M)$ is constant: 
\begin{equation}
\delta\left(\frac1{\alpha(0)}\right)=
-\frac 2{3\pi}\sum_fQ^2_f N_f\frac{\delta{m_f}}{m_f}
\end{equation}
In consequence, a compound variation
 by $-0.72\,10^{-5}\alpha^{-1}\,3\pi/2=-0.46$\%  in mass of (any of the)
three leptons $e,\mu,\tau$  and five quarks $u,d,s,c,b$
is sufficient to generate the observed effect. Restricting our attention
to the 2nd and 3rd generation and distributing the effect uniformly, we 
need less than -0.1\% variation in mass for each of the 5 fermions.

Within this chain of arguments we cannot expect that  
the variation of  the 
masses of second  and third  families is found in the 
conventional interactions, as these effects would also alter the
fundamental generation:
\begin{enumerate}
\item Masses, like coupling strengths, evolve with the interaction energy
scale, driven by variation of the coupling constant. It is 
customary to write, within QED, 
\begin{equation}
\frac M m \frac{dm}{dM}=\gamma(\alpha(M)).
\end{equation}
Given a universal 
function $\gamma(\alpha)$ for leptons, it is easy to see that the variation of all
lepton masses is exactly the same and thus the ratio of lepton masses
is  scale independent. For quark 
mass ratios, where  the scale evolution of strong 
interactions is  dominated by gluons, a similar argument
is  applicable in a very good approximation. Thus we cannot
introduce a variation of mass of 2nd and 3rd generation only 
by manipulating the scale running of masses.
\item Considering the case that the  masses of fermions are
derived from the coupling to the Higgs field, the mass ratios are 
ratios of coupling strengths. It is less than  obvious 
how a fraction of percent variability of mass  
can arise within the Higgs scheme solely for the 2nd and 3rd
generation.
\end{enumerate}

\vskip 0.3cm
{\bf Gravity for 2nd and 3rd generation:} To
explain the observed effect, we would like to see 
a variation of some of the masses, while keeping the fundamental 
generation masses constant. Thus by necessity we need an interaction\\
1) which breaks the fermion generation universality, and\\
2) which is sensitive to the conditions prevailing 
in the late Universe.

We propose a novel and speculative solution that each fermion
generation may have a different ratio of inertial to
gravitational mass, as is expressed by gravitational
coupling. We recall that the E\"otv\"os type  experiment 
establishes the ratio of 
gravitational and inertial mass solely for the ``stable'' 
first generation which surrounds us. 

To be very specific we propose a natural and very slight 
generalization of the  Einstein general relativity equations:
\begin{equation}\label{EinGen}
R_{\mu\nu}-\frac 1 2 g_{\mu\nu}R=\frac{\delta}{\delta g^{\mu\nu}}
\left( G_1I_1+G_2I_2+G_3I_3\right)
\end{equation}
Here, $R$ is  the Ricci curvature tensor, 
$G_i\equiv 8\pi g_i/c^2$ with $g_1=G$ the 
usual gravitational constant and $G_2, G_3$ are the gravitational 
constants of the 2nd and 3rd generation of particles, while $I_i$ is the
action of the generation $i$  derived from the study of inertial forces. 
Variation with respect to the metric $g^{\mu\nu}$ does not 
yield a global inertial energy momentum tensor $T_{\mu\nu}$, 
rather we find for each generation its $T_{\mu\nu}^i$-component
weighted with different strength of the non-universal gravitational interaction. Since each of the components, $T_{\mu\nu}^i$, has
the same transformation properties, the right hand side
of Eq.\,(\ref{EinGen}) transforms like the inertial
energy-momentum tensor.

We have not included gauge fields in the above. A priori
their universality poses a technical challenge
in our proposed approach. What needs to 
be derived to make our
approach correctly rooted in e.g. the brane theory is 
a brane configuration which maintains gauge universality, 
while yielding  gravity non-universality with 
respect to the different generations \cite{Die02}.
 However, within the  brane 
theory  more common is to find  a configuration leading to
gauge non-universal, gravity  universal theory \cite{Die02}. 

The inertial matter action $I_i$ is as given by 
Weyl, Fock and Iwanenko in late 20's:
\begin{equation}\label{FermiG}
I_i=\int d^nx\sqrt{-g}[\overline\Psi_i (\gamma^\mu iD_\mu -m_i)\Psi_i].
\end{equation}
Where $\gamma^\mu\gamma^\nu+\gamma^\nu\gamma^\mu=2g^{\mu\nu}$ defines
the Dirac-$\gamma$ general relativistic matrices and 
$D_\mu=\partial/\partial x^\mu+\Gamma_\mu$ with: 
$$\Gamma_\mu=\frac14\gamma_\nu\left(
{\partial\gamma^\nu\over\partial x^{\mu}}
+C^\nu_{\lambda\mu}\gamma^\lambda\right),
$$
and the Christoffel symbol is:
$$
C^\nu_{\lambda\mu}=\frac12 g^{\nu\beta}\left(
{\partial g_{\lambda\beta}\over \partial x^\mu}
+{\partial g_{\beta\mu}\over \partial x^\lambda}
-{\partial g_{\lambda\mu}\over \partial x^\beta}
\right).
$$

Cosmological solutions of field equations arising from Eq.\,(\ref{EinGen}) have 
the usual  structure. Presence of matter with different
gravitational couplings (e.g. cosmic background neutrinos) does 
not alter the structure of the Einstein equations. In the dust
model for the energy-momentum tensor the effective equations of state
are different, for the case of  mass less (neutrino)
gas the Einstein equations including the 2nd and 3rd generation
appear with effective degeneracy $g^{\rm eff}_{2,3}=g_{2,3}G_{2,3}/G_1$, there
seems to be more gravitating matter in the Universe, just as required.
Stellar objects containing   2nd and 3rd generation neutrinos would
still be defined by the asymptotic Schwarzschild mass, we could not
even tell which matter is within the star, short of locally probing 
the flavor and inertia of each component. Important for our later 
consideration of mass threshold is that the background Schwarzschild 
metric  attaches to the mass of a particle as 
\begin{equation}
m_i\to m_i\sqrt{1-\frac{2Mg_i}{r}}\to m_i\left(1-\frac{Mg_i}{r}\right).
\end{equation}
We thus see that the local gravitational potential attaches to the 
gravitational charge $g_i$ of particle `i'. Accordingly the 
mass defect is more significant for particles with greater coupling strength.

\vskip 0.3cm
{\bf Consequences of non-universal gravity:}
There is  lack of any direct experimental information on
the gravity action on 2nd and 3rd family. A very weak upper
limit of the strength of the interaction $G_{3,2}$ arises noting that
massive particles in these families should not be
beyond Planck mass limit, which requires 
$G_{3,2}/{G_1}< 10^{17}$--$10^{19}$. A stronger limit can be
derived from  loop contributions of virtual particles to physics
on Earth, which begins to play a noticeable 
role when $G_{3,2}/{G_1}\simeq 10^{10}$
An experimental study of muonium $\mu^+e^-$ fall in the gravitational
field of the Earth could reveal a better limit on $G_2$. During 
muons lifespan of 2$\mu s$ a particle falls 20\,\AA, thus 
a measurable effect requires  $G_2\gg G_1$. Short of 
approaching the Planck mass (see above) the enhanced
 gravitational interaction remains  weak. 

Turning to cosmological consequences we note that even if 
the  inertial masses
of neutrinos are at the (small) level of the mass differences seen  in 
neutrino oscillations, $m_i\simeq 10^{-2}$ eV \cite{Bar02}, a 1000 fold 
enhancement of  second and third neutrino generation gravitational coupling 
allows the two neutrinos in cosmological background to contribute
a gravitational mass-energy equivalent of nearly 10 eV$_G$
 each, and thus to describe the  `dark' gravitating mass of the Universe. 
The presence of more strongly gravitating component in the background of
our Universe is likely to help the dynamical evolution of density
fluctuations into the present day large inhomogeneities.  The
dynamics of the Universe expansion would need to be reexamined
in this new context.

One should not before a thorough exploration of the family 
specific gravitational interaction see the above factor $10^3$
as an upper limit on gravitational interaction enhancement. Namely, the
neutrino oscillation phenomenon combined with the large 
gravitational fields in early Universe can greatly alter this 
naive limit.

The required change of the fermion mass 
by -0.1--0.5\% within the nebulae which 
we are looking for is understood to be the gravitational 
mass defect. Our Sun  generates for normal 
gravity a mass defect $\delta m_1/m_1 =-0.3\, 10^{-5}$. Thus 
coincidental with the explanation of the dark matter, the non-universal  
gravity with $G_{3,2}/{G_1}\simeq 10^3$ yields the expected -0.3\% mass 
defect for the 2nd and 3rd generation fermions. 
Note that the sign of the
effect is correct, which aside of checking all equations 
can be seen as follows: background gravity reduces the effective
threshold for creation of a pair of particles, and this
reduction in effective mass allows the fine structure constant
to `run' faster and thus from a fixed unification value 
toward a smaller value, as is reported
experimentally to be present within the nebula. 

Once we interpret the variation of $\alpha$ as a
spatial effect, we could go back to check if it would
not be possible that all fermions are
subject to the mass defect. This would require that nebula
appear to have a gravitational interactions 
1000 stronger compared to  our sun also for regular matter. 
This seems not to be possible.  

Enhanced  gravitational coupling has the capacity to resolve some 
riddles in stellar structure. For example, the accretion 
of background neutrinos and resulting concentration of 
more strongly gravitating  neutrinos in 
large stellar objects  \cite{Bil01} 
releases greatly enhanced amount of gravitational energy as e.g. is
required to fuel the quasar phenomenon.  Since there are 
two different additional gravitational
couplings possible, and there is a finite supply of cosmic neutrinos 
it may be possible to explain both quasars and active galactic nuclei
in terms of accretion of neutrinos, and to limit the range
of cosmological time during which these phenomena occur. These
remarks are consistent with appearance of event horizons (black holes)
in such stellar objects, indeed the non-universal gravity of some of the
accreted (neutrino) particles would lead to faster formation of 
such a gravitational singularity. Naturally, this singularity is
the same for all test particles of any of the three generations.

\vskip 0.3cm
{\bf Final remarks:}
We conclude that the small deviation of $\alpha(0)$ 
from the established values could  be result of a somewhat 
larger (by factor $645=137\cdot 3\pi/2$) -0.46\% compound total
mass variation of  unstable second and third generation 
fermions. It seems  that a spatial rather than 
temporal variation of fermi masses, due to a gravitational
mass defect, is the source of this effect. The nebula 
which act as source of the absorption lines are required to
generate gravitational fields causing this
mass defects. An enhanced anomalous gravitational coupling of 
heavy fermi generations is our interpretational 
scheme. A factor 1000 enhancement in non-universal gravity
allows  interpretation
of dark matter as being due to  $\nu_\mu,\nu_\tau$ cosmic 
background, while  the gravitational mass defect 
of stable matter in absorbing nebula is at $10^{-5}$, just
a factor 5 greater than the mass defect of the sun.

\vskip 0.3cm
\noindent{\bf Acknowledgments:}
I thank Keith Dienes for stimulating discussions 
and valuable suggestions, and the CERN theory
division for hospitality while I was discovering this subject.
Work supported in part by a grant from the U.S. Department of
Energy,  DE-FG03-95ER40937.

\vskip 0.7cm

\end{document}